\documentstyle[twoside,11pt]{article}

\begin{document}

\renewcommand{\thefootnote}{\fnsymbol{footnote}}      

\thispagestyle{plain}
\setcounter{page}{1}

\vspace{6pc}

\vspace{3pc}

\vspace{2pc}
\begin{center}
CRITICAL FLUCTUATIONS IN THE BREAKDOWN
OF DISORDERED SYSTEMS
\end{center}
\vspace{0.27truein}
\centerline{Alberto Petri}
\hfill\break
\baselineskip=12.5pt
\centerline{\it Istituto di Acustica ``O.M. Corbino'', Consiglio Nazionale delle Ricerche}
\centerline{\it via del Fosso del Cavaliere 100, 00135 Rome, Italy}
\centerline{\it and}
\centerline{\it Dipartimento di Fisica e Istituto Nazionale di Fisica della Materia}
\centerline{\it Universit\`a La Sapienza, P.le A. Moro 2, 00185 Rome, Italy}
\vspace{0.36truein}
\abstract{In this paper some critical aspects of  the behaviour  of
breaking lattices subject to slow driving forces are briefly reviewed. 
In particular, 
fluctuations in the response to the variation of external parameters are discussed.
}{}{}
\vspace{0.78truein}
\baselineskip=14.5pt
\section{INTRODUCTION}
\noindent
\baselineskip=14.5pt
Antonio Coniglio's contribution to the undertstanding of
rupture phenomena in disordered  systems is twofold. On one hand he
has contributed indirectly to the field, by furnishing fundamental   
knowledge indispensable for dealing with dilute networks,  like e.g. 
the exponent for the scaling of the density of cutting bonds in a percolating lattice\cite{coniglio77}. 
On the other hand he has also pointed out important  properties of the response
of these systems,  such as  the
multifractality of currents 
in diluted\cite{dearcangelis85} and two component\cite{dearcangelis87} 
resistor networks. These studies have continued in 
the observation,  investigation and explanation of a large quantity of scaling 
properties in breaking lattices\cite{herrmannroux,chakrabeng}, 
and the great amount 
of work done in this field has shown that scale invariance is 
a fundamental property 
of rupture and breakdown phenomena of disordered 
systems\cite{herrmannroux,chakrabeng}. 
Indeed, disorder  plays a fundamental
role in generating this  kind of behaviour\cite{kahng88}.

Scale invariance
has been found  to be also a more general characteristic of
the dynamical response of these systems to the
variation of some external parameter, and  also a characteristic  of 
the fluctuations of their response. For this reason, 
the adjective ``critical'' is often used, 
in analogy with the fluctuations of  thermal systems at the 
critical point. In the following 
we shall mainly focus on this aspect of the breakdown phenomena  and will
consider in particular  those situations in which an  external perturbation
(driving) is  applied slowly with respect to the characteristic relaxation 
times of the system. After 
briefly reviewing some experimental results (Sec 2.), we shall describe 
how some of the observed 
features are reproduced by lattice models (Sec 3.), and some
current points of view on the subject are (Sec 4.).

\section{SOME EXPERIMENTAL FACTS}
\noindent

Initial discoveries about the critical response of a fracturing   
medium are probably due to Mogi\cite{mogi}. 
With the aim of verifying the validity of the Gutenberg Richter law
also at scales much smaller
than those involved in an earthquake, 
he designed and performed some
original experiments where  a controlled 
pressure was applied to disks made of a mixture of resin and hard grains. 
He observed that, by increasing  the pressure, 
elastic waves were released from some localised region within the 
sample. By
recording the maximum wave amplitude of each series Mogi was able to show that
the relation between the observed amplitudes and their frequency of occurrence
was of the algebraic type, the one known as the Ishimoto and Iida's law in the 
seismological field. 
These findings, besides to show in a quantitative way 
that earthquakes and fractures have  common features,  demonstrated 
the intrinsic critical nature and the importance of disorder in  the 
response of a medium to the external perturbations.

Laboratory experiments can  well reproduce not only the critical features
observed in distribustion of the energy released by earthquakes, but also  
those observed in the distribution of times. In 1968 Scholz succeeded in reproducing 
Omori's power law describing the number of aftershocks
observed in a time $t$ after a main event\cite{scholz68o}. By analysing 
the acoustic emission  from  a fracturing basalt rock
he also coonfirmed the validity of  the Gutenberg-Richter 
law\cite{scholz68gr}.  In subsequent investigations Hirata\cite{hirata87}
showed that Omori's law holds also for generic microfracturing processes,
at least
for times large enough after the main event. 

Thanks to the advances in technology and informatics,
experiments in this fields  continue to bring  new results. 
By collecting and analyzing acoustic  
emission signals from  concrete-like samples, we have 
shown\cite{petri94}  that  power laws not only   
describes  the frequency distribution of maximum amplitudes  
in the aftershock series, but also 
the amplitude distibution   
of the  entire time series,  
as well as the time lags distribution between consecutive meaningful 
events of acoustic emission. 
We have found in particular that the energy (proportional to the squared 
amplitude) is distributed according to 
$N(E) \simeq E^{-\delta}$ with 
$\delta$ around $1.3$, while
for  time lags one has $P(t) \simeq  t ^{-\zeta}$ with $\zeta$ approx $1.6$
\cite{vespi95}.
In addition, lacking of  characteristic scales in the fluctuations of response 
has been brought into evidence by measuring 
the power spectrum and the autocorrelation of the time series.
The power spectrum has been found of the  $1/f^\gamma$ form,  
with $\gamma\simeq 0.6$\cite{petri94,vespi95}. 

In more recent experiments Ciliberto and coworkers\cite{cilibe97,cilibe99} 
found $\delta \simeq 1.25$ for the energy scaling and conjectured
its universality,   
whereas $\zeta$ was found  to depend on the 
applied external stress. 
Acoustic emission recorded during pressurization of
spherical tanks yielded still more evidences of the critical nature of
breakdown phenomena and revealed  the presence of logarithmic oscillations
in the power laws\cite{sorn98}. Finally, Maes and coworkers\cite{maes98}
observed scale invariance in acoustic emission amplitudes, time lags, 
and spatial
distance between consecutive events also in a cellular glass, where
they found   $\zeta=1.3$, but $\delta=2$.

Another important evidence of the critical response of disordered
media concerns the roughness of the fracture surfaces\cite{mandel84}. 
After Mandelbrot and coworkers measured self-affinity properties in  
the fracture surface of some steels, similarity in the exponents 
($\approx 0.8$) characterizing many different materials
has been pointed out\cite{bouchaud90,maloy92}. Further experiments 
have  shown later that also  $0.5$ is observed\cite{milman93}. 
According to some  evidences   
lower values seem to characterize slow producing cracks and roughness at
small scales,
whereas higher values are related to large scales and should 
be associated with fast cracks\cite{mcanulty92,bouchaud97}.

\section{LATTICE MODELS}

Existence of critical fluctuations in the response of model systems 
was firstly observed~\cite{sorn89}~\cite{hansen92} 
in the fibre bundle model \cite{daniel45}, where an external applied stress 
is evenly shared by a stretched bundle of fibers.
The elastic modulus is the same for all the fibers,
but each one can stand a different, finite, amount  of 
stress. The system is initially unloaded, then stress is 
applied in order to break the weaker fiber of the bundle.
The excess stress is shared 
by the other fibres, that therefore become  more prone to 
break; whenver some fibre exceeds its own failure threshold it also breaks
and stress is redistributed again. When  no more fibres  
break
stress is increased again, up to break the weakest of the survived fibers, 
and so on until all the fibres of the bundle are broken.
Each breaking process is carried on at constant stress.
By exact calculation Hansen and Hammer were able to show 
that as the applied 
stress is increased from zero to the global failure value,
there is a probability $P(s)\propto s^{-\alpha}$ that $s$ fibers 
break in correspondance of the same value of applied stress.  
The exponent  was found to be
$\alpha=5/2$ and largely independent of the statistical distribution 
of fiber strenghts. 
Hansen and Hammer also observed  power law distributions for 
``avalanches' of broken fuses in the numerical simulation of a square 
resistor network where, 
in analogy with  he fiber bundle model, each fuse 
posseses the same conductance, but can stand 
different maximum current. When a fuse finds itself above its own 
threshold 
of rupture, it breaks and the excess current is shared 
by the other fuses. It may happen therefore that some other 
fuse burn in turn, and so on. 
They found numerically  $\alpha=2.7$ for this system,  
very close to $5/2$.

Presence of similar behaviour 
was later observed in system with vector elasticity by numerical 
simulations of the Born model on a triangular lattice\cite{caldarelli96}.
In this model sites interact via the potential
\begin{equation}
\label{eq2}
V_{ij}=(\alpha -\beta )\left[ \left( {\vec u}_i-{\vec u}_j\right)
\cdot {\vec r}_{ij}\right] ^2+\beta \left[ {\vec u}_i-{\vec u}_j\right]
^2,
\end{equation}
where ${\vec u}_i$ is the displacement vector of site $i$ from
equilibrium,  ${\vec r}_{ij}$ is the unit vector between the initial
equilibrium position of sites $i$
and $j$, and $\alpha $ and $\beta $ are force constants.
In the case of  Ref. \cite{caldarelli96}  
avalanches were  also 
triggered  by a corrosion mechanisms, according to which the 
bonds neighbouring  a broken bond are weakened.
Such a mechanism enhances the critical properties of the 
system\cite{caldarelli99a},
and the related exponent was found to be $\alpha=2.0$, the same found
in the experiments on the cellular glass\cite{maes98}.

The power law distributions mentioned above are computed by considering
all the avalanches occurring during the life of the system. 
One can also consider what happens for a given value of 
the external solicitation (stress, strain, current) 
$\sigma$ when averaging over many realizations. 
For the fibre bundle it has been shown 
that the following
scaling form  holds\cite{hansen92}:
\begin{equation}
\label{sacaling}
P(s,\sigma)= f(\frac{s}{s_0}) \ s^{-\tau}
\end{equation}
where 
\begin{equation}
s_0 \approx (\sigma_c-\sigma)^{-\kappa},
\end{equation}
and $\tau=1.5$ and $\kappa=1$;
$\sigma_c$ represents the critical value of stress at 
which the network definitively tears.
Thus, if one cumulates all the avalanches from the beginning,
$\sigma=0$, to the end of the system life, $\sigma=\sigma_c$,
one observes
\begin{equation}
P(s)=  s^{-\alpha}
\end{equation}
with $\alpha=\tau+1/\kappa=2.5$.
Zapperi at al.\cite{zapperi97} succeeded to show that within 
the effective medium 
approximation the fuse network obeys the same scaling of
the fibre bundle model,   
and that the ensemble averaged 
burst size $\langle s \rangle $ diverges at criticality as 
\begin{equation}
\langle s \rangle \approx (\sigma_c-\sigma)^{-\gamma}
\end{equation}
with $\gamma=1/2$. 
These results have been confirmed numerically both on  square
fuse networks\cite{zapperi97} 
and on elastic networks with central and bond bending potentials  
\cite{acharaya96}.

A way for quantifying the vicinity of a system to some
critical point
is to define a branching ratio $\rho$ for the process 
as the probability for  the breaking of a bond
to give rise to the breaking of another bond. In analogy with other
processes  \cite{prado98} at the critical point $\rho=1$.
We have evaluated $\rho$ for the Born potential  on a triangular 
lattice \cite{caldarelli99a} by 
averaging over many realizations the number 
of broken bonds at  fixed values
of the external driving (stress or strain)
$\langle s \rangle$. 
This quantity is related to $\rho$ by:
\begin{equation}
\rho=\frac{\langle s \rangle -1}{\langle s \rangle},
\end{equation}
and therefore
\begin{equation}
\rho\approx 1-(\sigma_c-\sigma)^{-\gamma}.
\end{equation}
Numerical simulations  show that
different types of driving produce
different behaviours,
as expected both from experiments and models\cite{herrmannroux}.
By computing  $\rho$  for the Born model we have shown in particular
that  breaking the bonds by driving with strain makes the system 
less 
critical than breaking them by driving with stress, and that
the presence of corrosion  mechanisms leads  closer 
to criticality.  

It must be stressed that the scale invariance
for  the breaking of bonds in model systems 
cannot be directly assimilated to that of acoustic emission. In fact, 
the latter
carries an energy content that is not included in the broken bonds counting.
Self-similarity of the energy bursts in lattice models only holds at 
comparable values of the external driving applied\cite{caldarelli96},  
or when special conditions are imposed to the system\cite{zapperi97a}.

Self affine properties of the fracture surface are rather well reproduced
by model simulations. In particular, numerical results suggest that
small self-affinity exponents (around $0.5$) 
are related to slowly moving cracks, whilst exponents close to $0.8$ 
are associated with
fast cracks \cite{chakra99,parisi00}. Some different exponents, like $0.42$  
have been also found for
slowly developping cracks in two dimensions \cite{nakano95}, but also if
simulations well reproduce generally the experimental observations,
there is no a comprehensive theory  for all these behaviours (for
a recent review on dynamic fractures see Ref.\cite{fineberg99}.

\section{POINTS OF VIEW}
\noindent
A  coherent description of breakdown phenomena and of their critical properties  is lacking.
One possibility is to relate them  to the 
Self Organized Criticality (SOC)\cite{btw86}. In this picture 
the SOC state corresponds to the final breakdown.
Another tempting  possibility is that of  exploiting the analogies with  critical
phenomena for setting up a thermodynamic-like description of the fracturing
process. Within this framework Zapperi et al.\cite{zapperi97} suggested 
to choose  the elastic moduli (or the conductance for the 
fuse network) as the order parameter, and drawed an analogy with
first order transitions, since the order parameter  suffers a 
jump at the final breakdown.  
Criticality, absent in usual first order transitions,
would originate in this case
from the presence of long range interactions; In different situations 
these have been 
shown to give rise to fractal fluctuations in the nucleation process\cite{klein90}.
The mean field picture, from which this interpretation is derived, does 
not describe all the features of the breaking process in euclidean lattices: 
pheraps the most relevant shortcoming is that the theory 
does not predict the scaling of the breaking 
stress distribution with the system size\cite{caldarelli99c}. 
Moreover, Sornette \cite{sorn97,sorn2} and da Silveira 
\cite{dasilv98,dasilv99} have
found that the choice of different disorder distributions  can select 
between  continuous and discontinuous transition, with a 
tricritical point separating
the two behaviours in the phase diagram of the system. 
Finally, continuous breakdown transition
can also be observed  when constant strain instead of constant stress driving
is applied to the lattice\cite{caldarelli99b}.

In conclusion we may say that in spite of the observation of critical fluctuations in many breakdown phenomena,
the physical mechanisms at their origin are still far from clear and
that it seems necessary to take into account both irreversibility 
and non self-averaging in order to come to a satisfactory explanation
of these complicated non-equilibrium processes.
 


\setcounter{footnote}{0}		
\renewcommand{\thefootnote}{\alph{footnote}}
\voffset=-1cm

\end{document}